\documentclass[11pt]{article}

\usepackage[final]{emnlp26}

\usepackage{times}
\usepackage{latexsym}
\usepackage[T1]{fontenc}
\usepackage[utf8]{inputenc}
\usepackage{microtype}
\usepackage{inconsolata}

\usepackage{graphicx}
\usepackage{booktabs}
\usepackage[normalem]{ulem}
\usepackage[most]{tcolorbox}
\usepackage{multirow}
\usepackage{soul}
\usepackage{pifont}
\usepackage{xcolor}
\usepackage[table]{xcolor}
\usepackage{eurosym}
\usepackage{tfrupee}
\usepackage{tikz}
\usepackage{subcaption}
\usepackage{footmisc}
\usepackage{wasysym}

\usetikzlibrary{patterns}
\tcbuselibrary{skins}
\tcbuselibrary{listings}

\definecolor{highlight}{HTML}{FAF7BD}
\colorlet{border}{brown!50}
\definecolor{green}{RGB}{60, 180, 120}
\colorlet{highlightgreen}{green!70}
\definecolor{red}{RGB}{180, 60, 60}
\colorlet{highlightred}{red!70}
\definecolor{forestred}{RGB}{242, 107, 0}
\definecolor{highInv}{RGB}{60, 180, 120}
\definecolor{lowInv}{RGB}{180, 60, 60} 

\newcommand{\rxmin}{1}
\newcommand{\rxmax}{5}
\newcommand{\likertchart}[3]{%
  \begin{tikzpicture}[xscale=0.32, baseline=-3pt]
    \draw[gray!50, very thin] (\rxmin, 0) -- (\rxmax, 0);
    \foreach \x in {1, 2, 3, 4, 5} {%
      \draw[gray, dashed, very thin] (\x, -0.07) -- (\x, 0.07);
      \node[font=\tiny, gray, anchor=north, inner sep=1pt] at (\x, -0.08) {\x};
    }
    \draw[forestred!50, line width=2.5pt, line cap=round] (#2, 0) -- (#3, 0);
    \fill[forestred!90!black] (#1, 0) circle (1.6pt);
  \end{tikzpicture}%
}

\newtcolorbox{promptbox}[1]{
  colback=white, colframe=black,
  boxrule=0.5pt, sharp corners,
  title=\textbf{#1},
  fonttitle=\small,
  coltitle=black, colbacktitle=white,
  titlerule=0.5pt,
  left=8pt, right=8pt, top=4pt, bottom=8pt,
}

\title{Expectations and Practices around AI Disclosure in CS Research}

\author{Arati Mohapatra \\
  Indian Institute of Science \\
  Bengaluru, KA, India \\
  \texttt{aratim@iisc.ac.in} \\\And
  Danish Pruthi \\
  Indian Institute of Science \\
  Bengaluru, KA, India \\
  \texttt{danishp@iisc.ac.in} \\}

\begin{document}
\maketitle

\begin{abstract}

As generative AI tools 
find increasing use 
in research workflows, 
ongoing debates on their impact, 
appropriateness 
and responsible use have led 
policymakers to enact
policies to disclose AI use at multiple publishing venues. 
However, are current AI disclosure policies 
and practices reflective of their purpose? 
In this work, we first investigate 
disclosure policies of top computer science venues
and find that despite their prevalence, they remain highly under-specified. 
Secondly, through a survey of computer science researchers (N=$109$), 
we characterize the necessity of disclosures 
across different research tasks and levels of human involvement.
We learn that researchers find disclosures 
most necessary 
for tasks involving research design, and 
for tasks when the human involvement is low. 
We also compile expectations that researchers 
have about the information 
to be conveyed in AI disclosure statements.
Lastly, through an analysis 
of $13867$ disclosure statements 
from EMNLP $2025$ and ICLR $2026$, 
we reveal a large disconnect between 
these expectations and AI disclosures in practice---a prime example being 
writing assistance which is deemed less necessary but is frequently disclosed. 
We conclude with recommendations to align AI disclosure policies and practices with expectations, suggesting a categorization of research tasks by perceived necessity and a boilerplate template capturing expected details.\footnote{Interactive survey results and code are available at \url{https://ai-disclosure.github.io/}}

\end{abstract}

\section{Introduction}

Accompanying the ongoing release of increasingly capable and efficient generative AI tools are discussions around the appropriateness of their use across professions \cite{inie_designing_2023, li_user_2024, butler_dear_2025}. The scientific community is similarly engaged in such a debate, and while researchers report improvements in quality and efficiency from the integration of generative AI tools into scientific workflows, concerns around their ethical use abound \cite{khalifa_using_2024}. Due to their ability to generate coherent natural language content on demand, generative AI tools may potentially shortcut intellectual contributions that were once considered indicators of rigorous human effort \cite{watermeyer_academics_2025}. Hence, to maintain the credibility, transparency, and reproducibility of research, strong arguments for disclosing generative AI use have emerged \cite{hosseini_ethics_2023, van_dis_chatgpt_2023, resnik_disclosing_2026}.

\begin{figure}[t]
  \centering
  \setlength{\fboxrule}{1.5pt}
  \setlength{\fboxsep}{0pt}
  \fbox{\includegraphics[width=0.98\columnwidth]{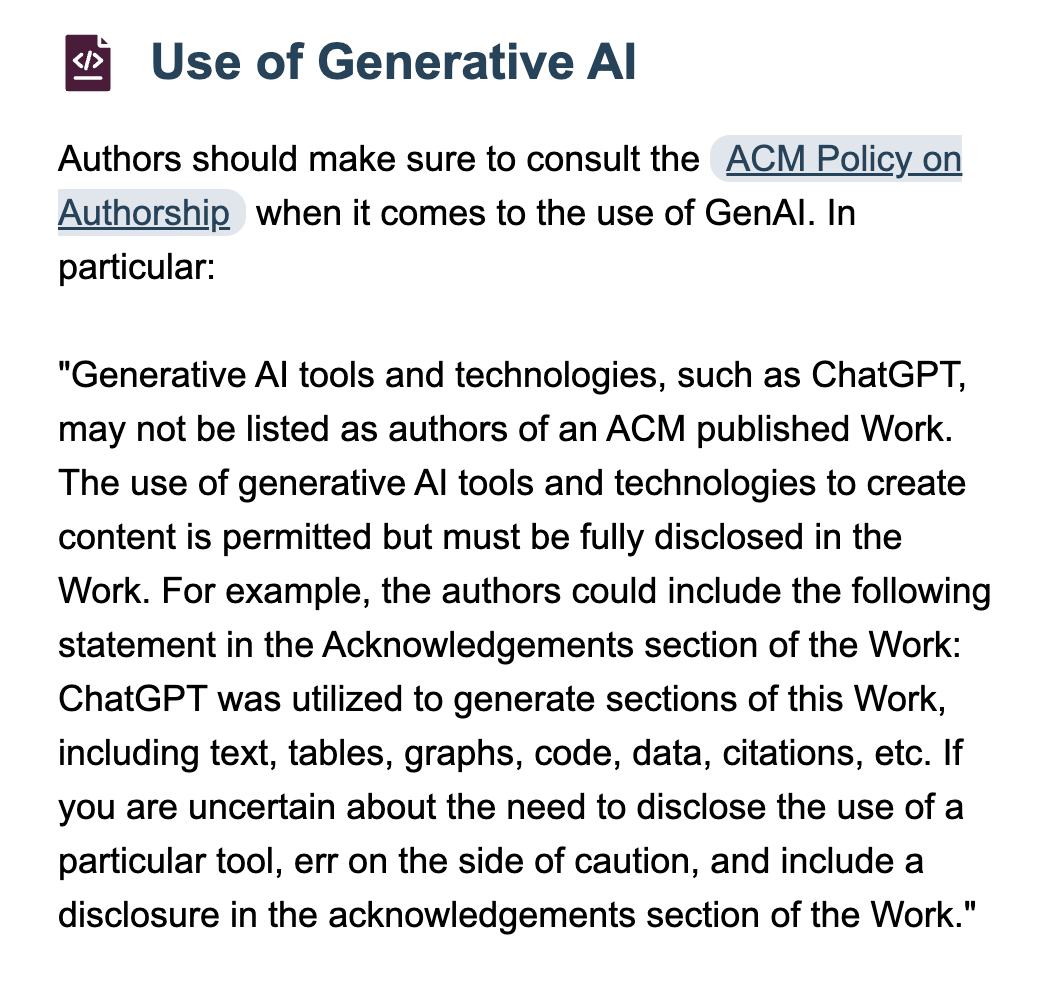}}
  \caption{ACM SIGCOMM's AI Disclosure policy.}
  \label{fig:policy}
\end{figure}

As a consequence, multiple publishing venues set up policies encouraging the inclusion of an AI disclosure statement in submitted research papers \cite{ganjavi_bibliometric_nodate,bhavsar_policies_2025}. The introduction of such policies was especially swift in the field of computer science, given the role of the community in developing these tools, and their relevance to computer science research \cite{chairs_acl_2023}. Relatedly, perceptions of quality and trust placed in research papers containing AI disclosures are mixed, perhaps due to shifting norms around generative AI use \cite{10.1145/3772318.3793386}. It is however evident that the presence of AI disclosures can shape the trust placed in scientific discourse. Yet, there have been minimal efforts to characterize the current state of AI disclosures in research. 

In this work, we first examine the prevalence and evolution of publishing venue policies that dictate the existence of AI disclosures in computer science research. We find that all major publishing societies (AAAI, ACL, ACM, and IEEE) and $35$ out of $65$ top computer science conferences have AI disclosure guidelines in place. However, we notice that they have stayed fairly constant since their introduction despite the rapidly evolving capabilities of AI tools, and provide limited guidance on \emph{when} generative AI use should be disclosed and \emph{what} specific details to include in a disclosure statement (see \autoref{fig:policy} for an example). 

To characterize current perspectives around the necessity and content of AI disclosures, we then conduct a survey of computer science researchers (N=$109$). Specifically, we ask the question: \emph{For what research tasks and levels of human involvement is it necessary to disclose generative AI use?} We observe that generally, participants find it more imperative to disclose generative AI use for research design and analysis than for writing. However, we also highlight the large variation in these opinions, which prevents us from recommending a few categories of tasks that warrant disclosure. Moreover, we show that the perceived necessity of disclosure depends on the level of human involvement during generative AI use, with higher levels of human involvement uniformly requiring less frequent disclosure. 

To understand whether current disclosure practices are reflective of these expectations, we analyze AI disclosure statements from EMNLP $2025$ and ICLR $2026$, two computer science conferences with established AI disclosure policies. We find that disclosures are included in $40$\% and $64$\% of research papers accepted at EMNLP $2025$ and submitted to ICLR $2026$ respectively. However, our analysis reveals a disconnect between expected disclosure behavior and how disclosing generative AI use plays out in practice. Specifically, while a declaration of responsibility for AI-assisted work is consistently expected, concerningly, $98$\% and $77$\% of AI disclosure statements in EMNLP and ICLR do not include this information. Moreover, the most frequently disclosed tasks are polishing text and editing code, for which disclosure is less expected. 

Overall, our findings underscore both the complexity of formulating precise AI disclosure policies and the misaligned nature of current disclosure practices. As initial steps, we recommend that policies move towards task-based guidelines, clarifying disclosure needs per research task rather than imposing a blanket requirement across all uses. Specifically, we suggest organizing research tasks into \emph{mandatory}, \emph{recommended}, and \emph{optional} categories based on the disclosure necessity scores assigned by the participants in our survey. We further recommend standardizing the format of disclosures, and design a boilerplate template that captures widely expected details (see \S\ref{sec:recommendations}). 

While we expect norms around AI disclosure to evolve organically, we hope this work will provide a starting point for the alignment of policies and practices with expectations, thus contributing to ongoing efforts to maintain credibility and transparency in research communication.

\section{Related Work}

\subsection{Generative AI in Research Workflows}
Given the recent interest in AI for Science, efforts are underway to develop AI tools to assist researchers with the aim of accelerating scientific discovery \cite{eger_transforming_2026}. Usage data from AllenAI's Asta, a research-specific generative AI service shows the prevalence of use for tasks from literature retrieval and idea generation to designing methods and writing \cite{haddad_understanding_2026}. The use of generative AI assistance at multiple stages spanning the research lifecycle is supported by researchers' self-reported Large Language Model (LLM) use \cite{liao2024llmsresearchtoolslarge}. Notably, perceptions of appropriateness of AI use vary by research tasks, which may in turn indicate differing necessity for AI disclosure \cite{kwon_is_2025}. Moreover, these norms themselves vary across different fields of research, with researchers in quantitative or experimental fields, such as computer science, demonstrating more openness towards both AI use and disclosure \cite{andersen_generative_2025}.

\subsection{AI Disclosure Policies}
Since the introduction of the first publicly available generative AI chatbots and tools in $2022$-$2023$, the policy landscape on AI use and disclosure has seen multiple shifts. Initial policies set up by publishers and journals across domains, while sufficiently prevalent, demonstrated considerable heterogeneity regarding the appropriateness of generative AI use and disclosure, particularly on when and how generative AI use should be disclosed \cite{ganjavi_bibliometric_nodate, bhavsar_policies_2025}. In computer science specifically, a temporal analysis of conferences policies from $2023$-$2025$ shows an uptick in the establishment of policies around generative AI use for authors, albeit with varying leniency and sanctions \cite{nahar_generative_2025}. Domain-agnostic attempts to standardize AI disclosure policies have recommended declaring generative AI assistance when used in an intentional and substantial manner, but have not yet reached consensus on the details to include in an AI disclosure statement, thus leaving significant room for author interpretation \cite{weaver_artificial_2024, resnik_disclosing_2026,bahammam_transparency_2025}. While these frameworks provide a valuable starting point for standardization, ensuring their representativeness of broader opinions of the research community remains essential. 

\subsection{Perceptions of AI Disclosures}
AI disclosures in research papers have been shown to impact perceptions of information quality and trust. A recent study shows that among abstracts disclosing differing levels of human and LLM involvement, human-written but LLM-edited abstracts receive the highest clarity ratings, whereas fully LLM-written abstracts score the least on both quality and trust scales, implying that the level of human involvement plays a role in influencing perceptions \cite{10.1145/3772318.3793386}. Moreover, this study shows that while the absence of an AI disclosure leads to speculations about the origins of the given text rather than quality evaluations, its presence puts these suspicions to rest, and increases trustworthiness. However, despite these recent findings demonstrating the positive effects of AI disclosure, hesitation towards full transparency may linger, due to a variety of reasons including how generative AI assistance was used, and social and personal factors \cite{fang_what_2025}. Evidence from a survey of $777$ researchers shows that self-decided appropriateness and the normalization of generative AI use are potential reasons for non-disclosure \cite{yusuf_understanding_2025}. Thus, in the absence of clear AI disclosure standards, non-disclosure may persist, with the potential to hamper transparency and trust in research \cite{bahammam_transparency_2025}.

\section{Research Questions} 
The review of related work highlights the prevalent use of generative AI assistance in research workflows, along with the attempts to introduce and standardize AI disclosure policies. This is especially imperative given the ability of declared AI assistance to impact the trust placed in research. Past work has also shown that these effects are closely entangled with the perceived appropriateness of generative AI use, that varies based on specific research tasks and levels of human involvement. In light of their potential impact, we attempt to characterize the current landscape of AI disclosures by addressing the following research questions: 

\begin{tcolorbox}[colback=white, colframe=border, boxrule=2pt, sharp corners]
\textbf{RQ1:} How prevalent are AI disclosure policies, and how have they evolved over time? 

\textbf{RQ2:} For what research tasks and levels of human involvement is it necessary to disclose generative AI use?

\textbf{RQ3:} How well do current AI disclosure practices align with reader expectations?

\end{tcolorbox}

We detail the procedures and results of the mixed-methods approach we took to answer these research questions in the following sections.

\section{AI Disclosure Policy Analysis}

\begin{tcolorbox}[colback=white, colframe=border, boxrule=2pt, sharp corners]

\textbf{RQ1:} How prevalent are AI disclosure policies, and how have they evolved over time?

\end{tcolorbox}

\subsection{Method and Procedure}
 To answer RQ1, we analyzed $65$ computer science conferences, examining the existence, origins and evolution of their AI disclosure policies. The $65$ conferences, spanning subdomains of Artificial Intelligence, Systems, Theory, and Interdisciplinary Areas were taken from the CSRankings website.\footnote{https://csrankings.org/} This list, developed in consultation with faculty and through community surveys, is representative of the top conferences in computer science. Since our aim was to characterize the current state of AI disclosure policies, we limited ourselves to an analysis of conference policies from $2025$-$2026$. We also analyzed the policies of AAAI, ACL, ACM, and IEEE, since many computer science conferences are affiliated with these societies, and hence may have contributed to shaping their current policies. We mapped each of the $65$ conferences to one or more of these society policies if they link to, mention, or verbatim quote a particular society policy. To trace the evolution of society policies since their origin, we used the Internet Archive's Wayback Machine,\footnote{https://web.archive.org/} which stores snapshots of web pages across time. We initially conducted these analyses from January $5$-$19$, $2026$, and reviewed and revised them from May $18$-$20$, $2026$. 

\subsection{Results}
Our analysis of $65$ computer science conferences shows that roughly half ($35$) of the conferences have set up AI disclosure policies. Moreover, we find that the major societies, AAAI, ACL, ACM, and IEEE, all have established AI disclosure policies: $29$ of the $35$ conferences with AI disclosure policies borrow closely from these society-level policies. Notably, $21$ conferences directly link to, mention, or verbatim quote the ACM policy on authorship, which includes guidelines on disclosing the use of generative AI. From our study of the evolution of society-level policies, we find that they have undergone only $1$ to $4$ changes since their first introduction in early $2023$, with revisions being minor, including restructuring existing content, adding reviewer guidelines or sanctions for prompt injections. We also note that the ACL and ACM policies include clear guidelines on the conditions necessitating disclosure of generative AI use based on either the novelty of generated text or the distinction between AI assistance for research versus writing. However, the AAAI and IEEE policies remain majorly open-ended, providing limited guidance on \emph{when} and \emph{where} to disclose generative AI use, only stating that any use of generative AI should be disclosed. Moreover, worryingly, none of the society-level policies provide specific guidance on what details are required to be mentioned while disclosing generative AI use.

\section{AI Disclosure Expectations Survey}
\label{sec:survey}

\begin{tcolorbox}[colback=white, colframe=border, boxrule=2pt, sharp corners]

\textbf{RQ2:} For what research tasks and levels of human involvement is it necessary to disclose generative AI use?

\end{tcolorbox}

\subsection{Method and Procedure}
\label{subsec:survey-method}

 To answer RQ2, we administered an online survey (on LimeSurvey) with $109$ respondents. Participants rated the necessity of AI disclosure for $21$ research tasks at $3$ levels of human involvement on a $5$-point Likert scale, where $1$ indicates that AI use need not be disclosed and $5$ implies that disclosure is always necessary. The chosen $21$ research tasks span the $5$ chronological phases central to the preparation of a research manuscript: idea generation, research design, data collection, data analysis, and writing and reporting. 

\begin{table}[t]
  \centering
  \small
  \begin{tabular}{p{0.2\columnwidth} p{0.675\columnwidth}}
    \toprule
    \textbf{Level} & \textbf{Description} \\
    \midrule
    \multirow{1}{=}{Assumed}
      & No explicit mention of human involvement. Participants \emph{assume} a certain baseline level of human involvement. \\
    \midrule
    \multirow{1}{=}{High}
      & When the author provides most of the contribution and initiative, while AI outputs are closely supervised and verified.  \\
    \midrule
    \multirow{1}{=}{Low}
      & When the AI assistant provides most of the contribution and initiative, and AI outputs are loosely supervised and verified by the author. \\
    \bottomrule
  \end{tabular}
  \caption{Descriptions of different levels of human involvement considered in the survey (See \S\ref{subsec:survey-method}).}
  \label{tab:human}
\end{table}

\begin{table}[t]
  \centering
  \small
  \begin{tabular}{p{0.2\columnwidth} p{0.7\columnwidth}}
    \toprule
    \textbf{Dimension} & \textbf{Variants} \\
    \midrule
    \multirow{7}{=}{Detail}
      & Task \\
      & Model name \\
      & Reason for AI use \\
      & Non-use of AI \\
      & Human oversight \\
      & Responsibility declaration \\
      & Purpose of disclosure \\
    \midrule
    \multirow{4}{=}{Length}
      & 1-2 sentences \\
      & A few sentences or a short paragraph \\
      & Multiple paragraphs \\
      & Length and detail proportionate to the extent of AI use \\
    \bottomrule
  \end{tabular}
  \caption{
    Dimensions along which survey participants 
    indicate their preferences for framing AI disclosures. %
  }
  \label{tab:framing}
\end{table}

\begin{figure*}[t]
  \centering
  \includegraphics[width=\textwidth]{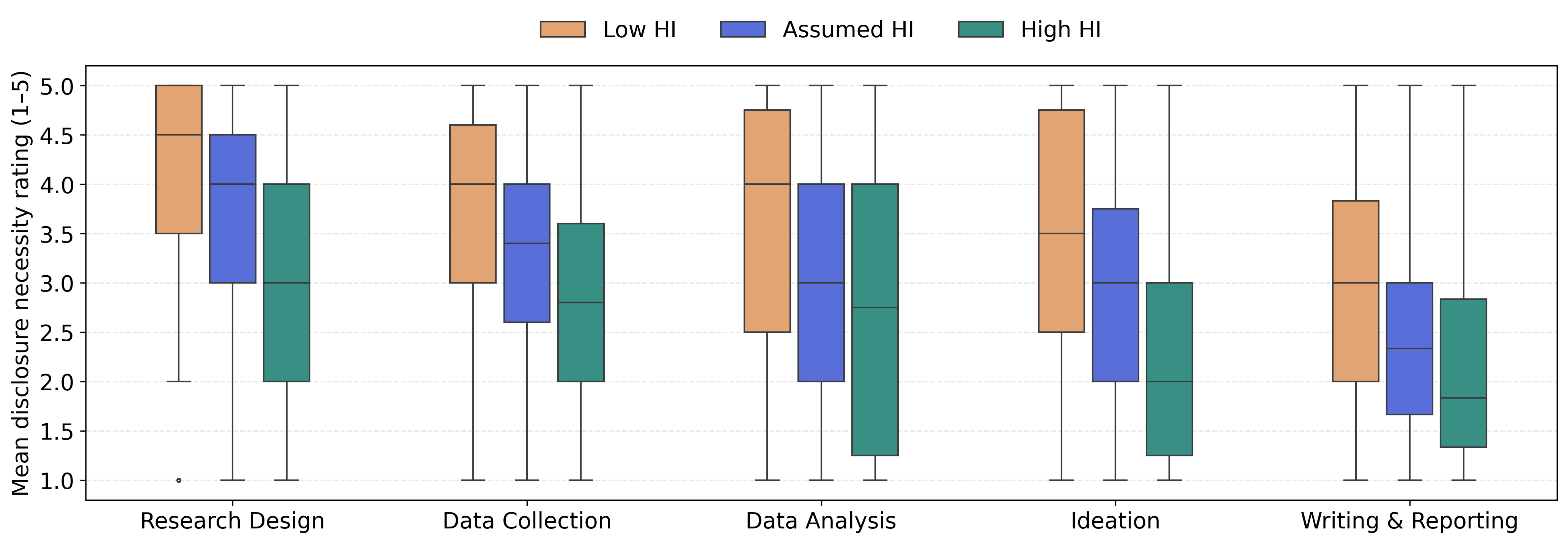}
  \caption{AI disclosure necessity ratings across research phases and human involvement (HI) conditions.}
  \label{fig:survey-phases-heatmap}
\end{figure*}

 The taxonomy of research tasks we used was adapted from \citet{andersen_generative_2025}, with minor changes to ensure suitability to our study and minimize cognitive load on participants. For instance, we removed all tasks related to peer review and research proposal writing, as these are unrelated to manuscript preparation, and also grouped together all coding tasks (for data analysis, statistical analysis, and simulations). We ensured that the final taxonomy is representative of tasks known to be assisted through generative AI tools \cite{liao2024llmsresearchtoolslarge,kapania_im_2025}. The research phase-wise categorization of tasks can be found in the Appendix, but individual tasks will also be listed in the subsequent sections, when we detail and discuss their necessity scores. 
 
 Given the diverse methods of using current generative AI tools, the appropriateness of use does not only hinge on \emph{what} AI assistance was used for, but also on \emph{how} it was used. Particularly, since computer science societies and conferences across the board, including ACL\footnote{\label{fn:aclsource}https://aclrollingreview.org/cfp\#authorship}, NeurIPS\footnote{https://neurips.cc/Conferences/2025/LLM} and RTAS\footnote{https://cmte.ieee.org/tcrts/conferences/transparency/}, stress the need for human intellectual contribution, integrity and verification, we try to understand whether the level of human involvement impacts AI disclosure necessity. The levels of human involvement used are described in \autoref{tab:human}.
 
To characterize expectations on the framing of an AI disclosure statement, we elicit preferences on details and length through multiple-choice questions.
The specific options, present in \autoref{tab:framing} were consolidated from existing conference guidelines and conceptual frameworks outlined in past work \cite{resnik_disclosing_2026,bahammam_transparency_2025,weaver_artificial_2024}. We empirically validate these options, by analyzing $110$ random AI disclosure statements from ICLR $2026$, which notably led to the inclusion of the \emph{Purpose of disclosure} and \emph{Non-use of AI} categories within the details dimension. 

To circulate our survey, we used a combination of purposive and snowball sampling through professional networks, university mailing lists, and departmental channels, targeting computer science researchers across AI, Systems, Theory and Interdisciplinary research. Participants span all $4$ subdomains, career stages ($<$$1$ to $>$$10$ years), and countries (e.g., US, Germany, India). The full survey instrument and participant demographics are included in the appendix. Participants were compensated with an Amazon Gift Card worth \$10/ \euro 10/ \rupee 500, depending on self-reported geographical location. The study received ethical approval from an institutional review board. 

To estimate differences in necessity between different research tasks and levels of human involvement, we fit a linear mixed effects model with necessity rating as the target variable, research task and human involvement as fixed effects, and participant-level random effects. Since norms around AI disclosure are still in flux, we expect substantive individual variability in respondents' ratings. Linear mixed-effects models, which account for such variability through a participant-level random intercept, can isolate the fixed effects of research task and human involvement from this between-participant noise. This model can be formulated as:
\begin{align*}
y_{\text{target}} &= \beta_0 + \beta_{\text{task}} \times x_{\text{task}} + \beta_{\text{human}} \times x_{\text{human}} \\
                  &\quad + u_{\text{part}} \times x_{\text{part}} + \epsilon,
\end{align*}

\noindent where $y_{\text{target}}$ corresponds to the user-provided necessity rating, $\beta_0$ is the fixed intercept, $x_{\text{task}}$ and $x_{\text{human}}$ are indicator variables for research task and level of human involvement, and $u_{\text{part}}$ is a participant-level random intercept.

\begin{table*}[ht]
  \centering
  \footnotesize
  \setlength{\tabcolsep}{3.55pt}
  \renewcommand{\arraystretch}{1.3}
  \begin{tabular}{l l l c}
    \toprule
    \textbf{Research Task}  & \textbf{Phase} & \textbf{Necessary?} & \textbf{Mean Rating} \\
    \midrule
    Generate synthetic data sets$^{*}$                              & Data Collection      & Usually     & \likertchart{4.02}{3.79}{4.25} \\
    Help develop theoretical models or conceptual frameworks$^{*}$  & Research Design      & Usually     & \likertchart{3.71}{3.47}{3.95} \\
    Propose new hypotheses$^{*}$                                    & Idea Generation      & Usually     & \likertchart{3.54}{3.27}{3.80} \\
    Help design research methodology or experiments$^{*}$           & Research Design      & Sometimes   & \likertchart{3.35}{3.11}{3.58} \\
    Translate a research paper into a different language$^{*}$      & Writing \& Reporting & Sometimes   & \likertchart{3.31}{3.05}{3.58} \\
    Clean and reformat dataset$^{*}$                                 & Data Collection      & Sometimes   & \likertchart{3.24}{2.99}{3.49} \\
    Help pattern recognition in data$^{*}$                           & Data Analysis        & Sometimes   & \likertchart{3.21}{2.95}{3.47} \\
    Support qualitative and thematic data analysis$^{*}$              & Data Analysis        & Sometimes   & \likertchart{3.17}{2.90}{3.45} \\
    Formulate questions for surveys or interviews                     & Data Collection      & Sometimes   & \likertchart{3.09}{2.83}{3.34} \\
    Create or modify scientific figures or images                     & Data Analysis        & Sometimes   & \likertchart{3.00}{2.73}{3.27} \\
    Suggest experimental parameters                & Data Collection      & Sometimes   & \likertchart{2.96}{2.72}{3.21} \\
    Create or edit software code                     & Data Analysis        & Sometimes   & \likertchart{2.96}{2.70}{3.21} \\
    Draft parts of a research paper                                   & Writing \& Reporting & Sometimes   & \likertchart{2.95}{2.70}{3.21} \\
    Transcribe recordings of research material                        & Data Collection      & Sometimes   & \likertchart{2.93}{2.66}{3.20} \\
    Summarize or analyse existing literature                          & Idea Generation      & Sometimes   & \likertchart{2.85}{2.60}{3.10} \\
    Discover research topics or identify gaps$^{*}$                 & Idea Generation      & Sometimes   & \likertchart{2.58}{2.33}{2.83} \\
    Edit a research paper to improve readability$^{*}$              & Writing \& Reporting & Rarely      & \likertchart{2.47}{2.24}{2.71} \\
    Identify relevant literature$^{*}$                              & Idea Generation      & Rarely      & \likertchart{2.46}{2.23}{2.68} \\
    Format references$^{*}$                                         & Writing \& Reporting & Rarely      & \likertchart{2.28}{2.04}{2.52} \\
    Suggest a structure for a research paper$^{*}$                  & Writing \& Reporting & Rarely      & \likertchart{1.99}{1.78}{2.20} \\
    Propose a title or keywords for a research paper$^{*}$          & Writing \& Reporting & Rarely      & \likertchart{1.95}{1.74}{2.16} \\
    \bottomrule
  \end{tabular}
\caption{Research tasks ordered by mean disclosure-necessity rating. The third column reports the necessity of disclosure, by identifying the nearest 
option in the Likert scale, and the Mean Rating column visualizes each task's mean and 95\% confidence interval. Significance markers ($^{*}p<.05$) indicate differences from the median-rated task (\emph{Suggest experimental parameters}) in a linear mixed-effects model.}
\label{tab:tasks-results}
\end{table*}

\subsection{Results}
\label{subsec:survey-results}

Overall, we observe that participants \emph{find disclosing generative AI use moderately necessary}, with a mean necessity rating of $2.95$ ($95$\% confidence interval: [$2.78$, $3.13$]). Through a subsequent research phase-wise analysis, we observe that AI disclosure is the most necessary for tasks in the research design phase, and the least necessary for tasks in the writing and reporting phase (see \autoref{fig:survey-phases-heatmap}). Contrary to what we had expected given the emphasis on novel ideas in research, we find that disclosing generative AI assistance for tasks from the idea generation phase is considered relatively less necessary, with a mean rating of only $2.86$. 

We also observe a surprisingly high amount of variability in the perceived disclosure necessity across all phases and levels of human involvement. This effect could potentially be explained by individual tasks within the same phase receiving wildly different ratings. For instance, within the idea generation phase, \emph{Proposing new hypotheses} (mean=$3.54$) and \emph{Identifying relevant literature} (mean=$2.46$) differ by more than an entire point on the Likert scale, and fall on opposite sides of the median-rated task. Tasks within the data collection phase, specifically \emph{Generating synthetic data sets} (mean=$4.02$) and \emph{Transcribing recordings of research material} (mean=$2.93$), exhibit similar differences in ratings. Task-wise necessity 
ratings are available in \autoref{tab:tasks-results}, where we also observe that most writing and reporting tasks are clustered together at the lower end of the necessity scale.

Interestingly, across all research phases, shifts from the \emph{assumed} level of human involvement exhibit a consistent trend: \emph{low human involvement raises disclosure expectations}, ($\beta$=$0.49$, $95$\% CI=[$0.42$, $0.55$], $p<0.001$) while \emph{high human involvement causes drops in the perceived necessity of disclosure} ($\beta$=$-0.44$, $95$\% CI=[$-0.51$, $-0.38$], $p<0.001$). The shifts in either direction, are however not always similar. 
For instance, the median necessity ratings for the low and assumed human involvement conditions in the data analysis phase are $1$ point apart, while this difference for the assumed and high human involvement conditions is less than $0.5$. The direction of asymmetry is reversed in the idea generation and research design phases, with a $1$-point drop in necessity ratings from the assumed to the high human involvement conditions, but only a $0.5$-point increase from the assumed level of human involvement to the low human involvement condition. This indicates that the loss of human agency in certain phases (e.g., in data analysis) drives disclosure necessity up, perhaps due to the potential errors that could arise in this situation, while the assurance of retained human agency for certain phases (e.g., in idea generation) lowers the need for disclosure. 

Our findings complicate recommendations that outline the necessity of disclosing generative AI use based on a categorization of tasks, given the high variability in necessity ratings per phase. \citet{resnik_disclosing_2026} advocate for disclosure when AI-assistance either directly impacts research results or is used to generate or analyze content, data or images. While research tasks fitting this description are generally assigned higher necessity ratings by our participants, we highlight the observed variability, as illustrated by the difference in perceived necessity of disclosure for \emph{Generating synthetic datasets} (mean=$4.02$) and \emph{Creating or modifying scientific figures or images} (mean=$3.00$). Similarly, we show that while \emph{Creating or editing software code} is stressed upon by ACM's guidelines,\footnote{https://www.acm.org/publications/policies/new-acm-policy-on-authorship} this is considered less necessary by our participants (mean=$2.96$). Further, ACL's guidelines\footref{fn:aclsource} that necessitate AI disclosure depending on the novelty of text or ideas produced, evoke similar mismatches, particularly for idea generation tasks, where participants find AI disclosure to be relatively less needed. Lastly, we highlight tasks that exhibit high perceived necessity, but cannot be neatly compartmentalized based on current policy guidelines: \emph{Translating a research paper} (mean=$3.31$) and \emph{Cleaning and reformatting data sets} (mean=$3.24$), both tasks that could introduce subtle erroneous artifacts if AI-assisted.

\section{AI Disclosure Practices Audit}

\begin{tcolorbox}[colback=white, colframe=border, boxrule=2pt, sharp corners]

\textbf{RQ3:} How well do current AI disclosure practices align with reader expectations?

\end{tcolorbox}

\subsection{Method and Procedure}
To answer RQ3, we extracted AI disclosure statements from $19525$ research papers submitted to ICLR $2026$ and $3216$ research papers accepted to EMNLP $2025$ Main and Findings, and analyzed the prevalence of tasks and details being disclosed. We chose to focus on these $2$ specific conferences due to their established AI disclosure policies, their recency, and the open availability of research papers. Interestingly, the two conferences present a key policy difference: ICLR $2026$ encouraged the inclusion of AI disclosure statements in a separate section within the submitted manuscript, while EMNLP $2025$ allowed for generative AI use to be disclosed either in the main text or in the author submission checklist. This allows us to also comment on differing disclosure behavior due to differences in policy design. 

We accessed accepted, withdrawn, rejected, and desk-rejected ICLR research papers through the OpenReview API, and EMNLP Main and Findings submission checklists and research papers from the ACL Anthology. We then used Gemini 2.5 Flash to evaluate the presence of and extract AI disclosure statements in research papers or submission checklists. We also used Gemini 2.5 Flash to annotate extracted AI disclosures with the details, tasks, and length present, in accordance with the taxonomies presented in \autoref{tab:framing} and \autoref{tab:tasks-results}. For extraction, Gemini achieved $100$\% F1 against human annotations on $100$ papers, while on $100$ disclosures annotated by three researchers, Gemini scored average micro F1 of $96.5$\% (details) and $90.6$\% (tasks). While these scores are not perfect, we believe they are satisfactory for our extraction and annotation purposes.

\subsection{Results}

\begin{figure}[t]
  \includegraphics[width=\columnwidth]{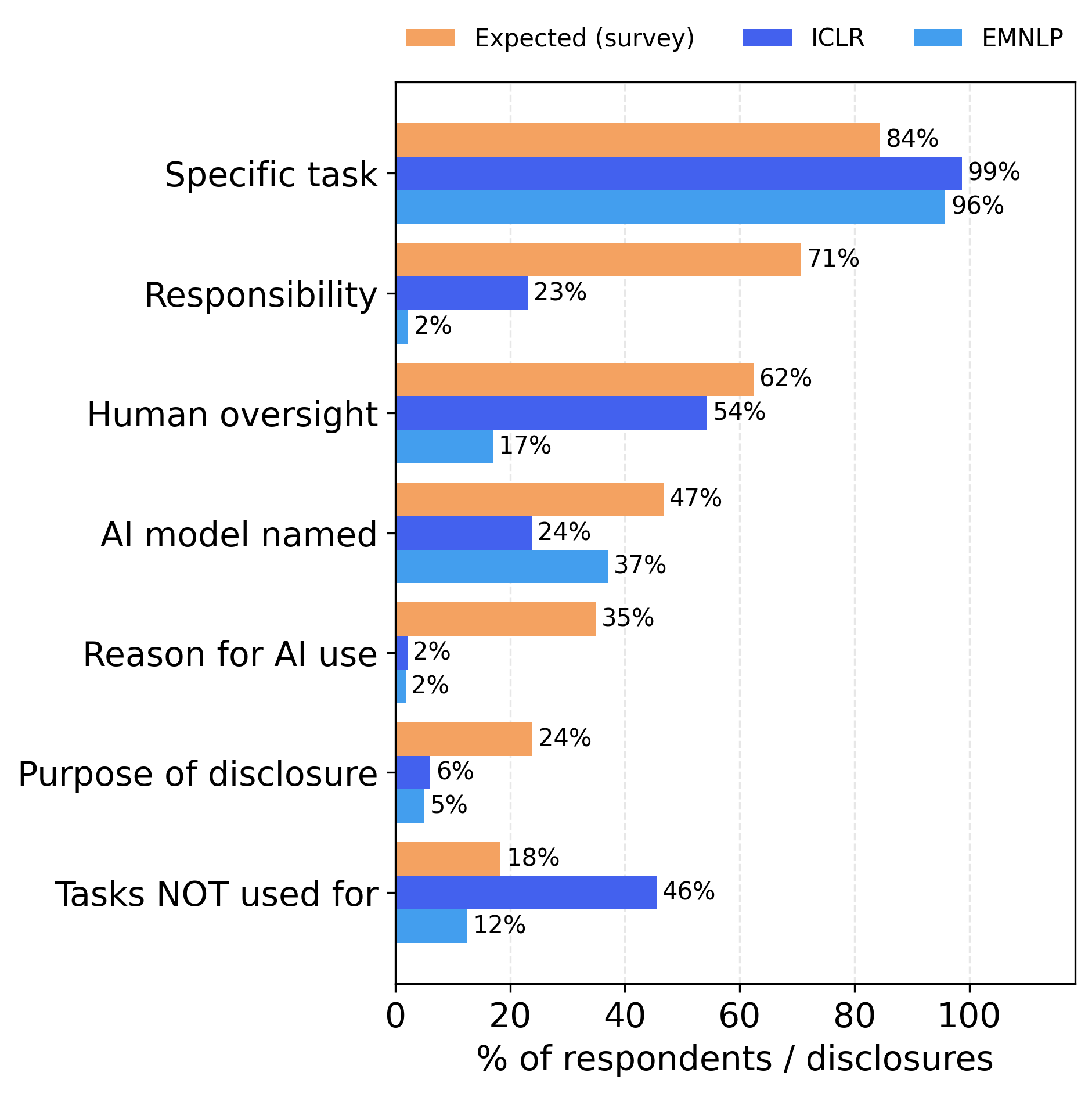}
  \caption{Expected vs. observed coverage of disclosure details, comparing survey expectations against ICLR and EMNLP practice.}
  \label{fig:details-disclosed}
\end{figure}

Despite the lack of strict enforcement of AI disclosure policies,
our analysis of $12577$ extracted ICLR disclosures and $1290$ extracted EMNLP disclosures reveals a moderately high prevalence of AI disclosures, with $64$\% of research papers submitted to ICLR and $40$\% of EMNLP Main and Findings research papers containing disclosure statements. We observe that tasks with lower perceived necessity are disclosed the most frequently: \emph{Editing a research paper} ($96.5$\% in ICLR and $81.4$\% in EMNLP) and \emph{Creating or editing software code} ($16.1$\% in ICLR and $23.3$\% in EMNLP). On the contrary, only a small fraction of disclosures exist for higher-rated tasks such as \emph{Generating synthetic data sets} ($2.1$\% in ICLR and $2.5$\% in EMNLP) and \emph{Translating a research paper} ($1.6$\% in ICLR and $2.0$\% in EMNLP). This gap may exist because generative AI use for tasks like polishing text or editing code is seen as more appropriate, and is therefore more common and more frequently disclosed.

\begin{figure}[t]
  \centering
  \begin{tcolorbox}[
    colback=orange!7, colframe=orange!60!black,
    boxrule=0.8pt, arc=2pt,
    title=\footnotesize\textit{Disclosure found verbatim in 95 ICLR 2026 submissions},
    fonttitle=\bfseries,
    colbacktitle=orange!25, coltitle=black,
    attach boxed title to top left={xshift=4pt, yshift=-4pt},
    boxed title style={sharp corners, boxrule=0.8pt, colframe=yellow!50!black},
    left=6pt, right=6pt, top=10pt, bottom=6pt,
  ]
  \small\itshape
  Large Language Models (LLMs) were used to aid in the writing and polishing of the manuscript. Specifically, we used an LLM to assist in refining the language, improving readability, and ensuring clarity in various sections of the paper. The model helped with tasks such as sentence rephrasing, grammar checking, and enhancing the overall flow of the text.

  \vspace{0.5em}
  It is important to note that the LLM was not involved in the ideation, research methodology, or experimental design. All research concepts, ideas, and analyses were developed and conducted by the authors. The contributions of the LLM were solely focused on improving the linguistic quality of the paper, with no involvement in the scientific content or data analysis.

  \vspace{0.5em}
  The authors take full responsibility for the content of the manuscript, including any text generated or polished by the LLM. We have ensured that the LLM-generated text adheres to ethical guidelines and does not contribute to plagiarism or scientific misconduct.
  \end{tcolorbox}
  \caption{An identical AI disclosure statement appearing in 95 ICLR 2026 submissions.}
  \label{fig:duplicate-disclosure}
\end{figure}

 Despite the structural difference between ICLR's free-text disclosure statements and EMNLP's checklist responses, we observe similar disclosure behavior in both venues, as shown in \autoref{fig:details-disclosed}. However, concerningly, while $71$\% of participants expect disclosures to contain a statement of responsibility from the authors, only $2$\% and $23$\% of EMNLP and ICLR disclosures respectively include such a declaration. It is also interesting to note that almost $50$\% of ICLR disclosures disclose the tasks generative AI was not used for, but this detail is considered the least necessary by participants. While authors may understandably include such information to emphasize the scope of AI assistance and their own contributions, perhaps this information is better included while expanding on the extent of human oversight, which is more frequently expected ($62$\%). Moreover, surprisingly, less than $50$\% of participants exhibit a preference for the inclusion of the specific AI model used, perhaps due to the limited number of commercial models in widespread use, and their roughly equivalent capabilities. 
 
 We also find that while the general expectation for the length of an AI disclosure statement is \emph{A few sentences or one short paragraph}, most disclosures in practice tend to be $1$-$2$ sentences long. This prevalence of shorter disclosure statements ($91$\%) is especially pronounced in EMNLP. This is in contrast to ICLR, where a sizeable chunk of disclosures ($41$\%) are a few sentences long.

\definecolor{templatebg}{RGB}{232, 245, 233}    %
\definecolor{templateborder}{RGB}{56, 142, 60}  %

\begin{figure}[t]
  \centering
  \begin{tcolorbox}[
    colback=templatebg, colframe=templateborder,
    boxrule=0.8pt, arc=2pt,
    title=\footnotesize\textbf{Proposed AI Disclosure Template},
    fonttitle=\bfseries,
    colbacktitle=templateborder!20, coltitle=black,
    attach boxed title to top left={xshift=4pt, yshift=-4pt},
    boxed title style={sharp corners, boxrule=0.8pt, colframe=templateborder},
    left=6pt, right=6pt, top=10pt, bottom=6pt,
  ]
  \small\itshape
  In this work, we used generative AI tools for \textless some mandatory disclosure tasks\textgreater. We have not used generative AI tools for \textless other mandatory disclosure tasks\textgreater, and \textless the rest of mandatory disclosure tasks\textgreater{} are not applicable to this work. Additionally, we used generative AI tools for \textless some recommended disclosure tasks\textgreater{} [and optionally for \textless some optional disclosure tasks\textgreater]. 
  
  \vspace{0.5em}
  We have reviewed all AI-assisted work [Elaborate. For example, “We checked LLM-generated research ideas for potential plagiarism through a manual literature survey”, “LLM-generated code was verified and tested for correctness by 2 authors”, etc.]. We take responsibility for the final content of this work, including text/claims/artifacts produced with the aid of generative AI.
  \end{tcolorbox}
  \caption{A structured AI disclosure template, designed to capture details that survey respondents prioritize.}
  \label{fig:disclosure-template}
\end{figure}

Alarmingly, during our analysis, \emph{we uncover the existence of multiple duplicate disclosures}. One such disclosure statement (c.f., \autoref{fig:duplicate-disclosure}) was found verbatim in $95$ unique ICLR submissions.\footnote{We initially suspected that this could be due to shared authors, but that was not the case. However, state-of-the-art AI text detector, Pangram, flags the statement as AI-generated.} While the statement itself is comprehensive and includes all expected details, its recurrence across unrelated papers suggests that the disclosure is not specific to any individual submission's actual AI use. Given the difficulty of enforcing disclosure policies, this highlights the potential issue of performative disclosures, intended to comply with policies rather than to inform.

\section{Recommendations}
\label{sec:recommendations}

\begin{table}[t]
  \centering
  \footnotesize
  \renewcommand{\arraystretch}{1.1}
  \begin{tabular}{@{}>{\raggedright\arraybackslash}p{0.94\columnwidth}@{}}
    \toprule
    \textbf{Research Tasks by Disclosure Level} \\
    \midrule
    {\textbf{Mandatory}} \textnormal{(rating $\geq 3.5$)} \\
    \addlinespace[2pt]
    Generate synthetic data sets \\
    Develop theoretical models or frameworks \\
    Propose new hypotheses \\
    \midrule
    {\textbf{Recommended}} \textnormal{($2.5 \leq$ rating $< 3.5$)} \\
    \addlinespace[2pt]
    Design research methodology or experiments \\
    Translate a research paper \\
    Clean and reformat a dataset \\
    Pattern recognition in data \\
    Qualitative and thematic data analysis \\
    Formulate survey or interview questions \\
    Create or modify scientific figures \\
    Suggest experimental parameters \\
    Create or edit software code \\
    Draft parts of a research paper \\
    Transcribe research recordings \\
    Summarize or analyze existing literature \\
    Discover research topics or gaps \\
    \midrule
    {\textbf{Optional}} \textnormal{(rating $< 2.5$)} \\
    \addlinespace[2pt]
    Edit a paper for readability or language \\
    Identify relevant literature \\
    Format references \\
    Suggest a structure for a paper \\
    Propose a title or keywords \\
    \bottomrule
  \end{tabular}
  \caption{Suggested disclosure levels for research tasks, based on ratings from survey participants.}
  \label{tab:task-lists}
\end{table}

Informed by our findings, we recommend that publishing venues specify \emph{when} disclosure is needed, and standardize \emph{what} a disclosure should contain.

\paragraph{Task-based AI disclosure policies} The results of our survey show that disclosing AI use is not considered equally necessary across research tasks, and we also note that existing policy categories, at their current granularity, do not reflect these varying expectations. While we acknowledge that policies around AI disclosure are hard to formulate precisely, we recommend that wherever possible, policies specify disclosure requirements at the level of individual tasks instead of broad rules. Such policies could include a list of tasks at differing levels of disclosure necessity: \emph{mandatory disclosure} for tasks where disclosure is widely expected (e.g., \emph{Proposing new hypotheses}), \emph{recommended disclosure} for tasks where expectations are more diffuse (e.g., \emph{Creating or editing software code}), and \emph{optional disclosure} for tasks where disclosure is deemed less necessary (e.g., \emph{Editing or polishing text}). We encourage policymakers at publishing venues to reproduce our survey, and use the resulting task-wise necessity ratings to inform the level of disclosure appropriate for different research tasks, for example, classifying tasks scoring $\geq 3.5$ as \emph{mandatory}, $2.5$–$3.5$ as \emph{recommended}, and the rest as \emph{optional}.

\paragraph{Standardized AI disclosure formats} Our analysis of AI disclosures in practice shows that current disclosure statements are rarely aligned with expectations, and may be performative. We speculate that this is in part due to the absence of a common format. We construct a boilerplate template from the details expected by over half of our survey participants, and recommend that publishing venues include it as an example in their policies and embed it directly within conference-specific \LaTeX{} files. We also suggest this boilerplate template include the tasks requiring disclosure, as outlined in \S\ref{subsec:survey-results}. We provide an example boilerplate disclosure statement in \autoref{fig:disclosure-template}, with task lists omitted for brevity, but included in \autoref{tab:task-lists}. A disclosure within the paper primarily serves readers; we additionally recommend a dedicated field in author submission checklists for disclosing \emph{mandatory} tasks, which would better serve reviewers and area chairs.

\section{Conclusion}
In this work, we examined AI disclosure in computer science research through a policy analysis of $65$ conferences, a survey of $109$ researchers, and an audit of $13867$ disclosure statements. We find that high variability in the perceived disclosure necessity across research tasks complicates the formulation of precise disclosure policies, and that current author practices diverge substantially from these expectations. Our findings offer empirical grounding to inform ongoing efforts to standardize and maintain transparency in research communication.

\section*{Limitations}

\paragraph{Self-selection bias} Our survey relied on convenience-based recruitment,  and participants who chose to respond likely already held opinions about AI disclosure. Expectations from a broader, less self-selected population may be weaker or more diffuse than those we report. Future work could mitigate this through stratified sampling across computer science subdomains or randomized recruitment via venue mailing lists. 

\paragraph{Scope to CS Research}
Our findings are limited to computer science research, where AI tools are arguably most familiar and disclosure norms are evolving most rapidly. While this focus is deliberate given that disclosure expectations likely differ across disciplines, extending this characterization to other research communities is a natural direction for future work.

\paragraph{LLM Annotation Reliability} We also make use of an LLM judge (Gemini 2.5 Flash) to annotate disclosure statements. Although we validated these annotations against human annotators with high agreement (average micro F1 of $96.5$\% (details) and $90.6$\% (tasks)), residual errors are possible, particularly for rare details or task categories with limited validation support.

\section*{Ethical Considerations}

This work characterizes AI disclosure expectations through a survey with human participants. Informed consent was explicitly confirmed at the beginning of the survey, and respondents were given the option to stop participation at any point. The survey was completely anonymous, and providing demographic details was optional. Participants who completed the survey were compensated with Amazon Gift cards worth \$10/ \euro 10/ \rupee 500, depending on self-reported geographical location. We only collected contact details (such as e-mail) to dispense the gift cards, and this information was deleted immediately after confirmation of receipt of compensation. As these details were obtained through a separate form, linking contact details to survey responses (thus risking deanonymization), was not possible. This study received approval from an institutional review board. 

Our work further recommends policy changes, and while these are made in the spirit of improving transparency and are grounded in community opinions, AI disclosures carry broader consequencess. For instance, researchers who disclose legitimate uses of AI assistance such as translation may inadvertently reveal that they are non-native English speakers, an attribute that, once inferred during peer review, risks reinforcing existing biases against them. To amend this, we suggest masking language support disclosures during the review period. More broadly, AI disclosures may invite negative perceptions of the work. While current norms around the appropriateness of AI use are still in flux, this cannot be fully eradicated. We nonetheless believe that standardized disclosure formats and processes from policymakers, along with authors' thoughtful efforts to emphasize the extent of their oversight and responsibility will help. On the other hand, disclosures may also occasionally benefit authors: because they are made a priori, they can serve as evidence for the actual extent of AI use when work is incorrectly flagged for violating AI use policies by venues relying on (potentially inaccurate) AI text detectors.

\section*{Acknowledgments}

We are grateful to all participants who took part in our survey and (whenever possible) forwarded it. We thank Rounak Saha and Rose Sathyanathan for annotations to validate the LLM judge used to label AI disclosure statements. We also thank Shobini NS for providing feedback on initial versions of the survey, and Kirti Bhagat for thoughtful discussions on interpreting results. We are also grateful to Vani A for administrative support. This work was supported in part by Schmidt Sciences, and additionally, DP is grateful to Google, Microsoft Research, and the Indian Institute of Science for supporting his group's research. 

\paragraph{AI Disclosure} In this work, we used generative AI tools (Claude Code) to create and edit software code, specifically for parallelizing the AI disclosure statement extraction pipeline. All generated code was manually verified for correctness. We have not used generative AI tools for generating synthetic data sets, developing theoretical models/frameworks, or proposing new hypotheses. We take responsibility for the final content of this work, including the artifacts produced with the aid of generative AI.

\bibliography{emnlp26}

\appendix

\section{Appendix}
\label{sec:appendix}

The contents of the Appendix are organized as follows: 
\autoref{tab:conferences} outlines the prevalence of AI disclosure policies across computer science conferences, grouped by subdomain. \autoref{tab:policies} shows the mapping of conferences with established AI disclosure policies to corresponding societies, along with the number of revisions the society policies have undergone since their introduction. \autoref{tab:tasks} details the taxonomy of research tasks we used, categorized by research phase. \autoref{tab:details} contains examples of extracted disclosure statements, grouped by type of detail. Tables \ref{tab:survey-welcome}--\ref{tab:survey-demographics-text} include the full survey instrument. \autoref{tab:survey-demographics} contains survey participant demographic information. Tables \ref{tab:extraction-prompt}--\ref{tab:emnlp-extraction-prompt} detail the prompts used for the extraction of AI disclosure statements. \autoref{tab:annotation-validation} and \autoref{tab:iaa} show validation results of Gemini 2.5 Flash for annotating extracted disclosures. Tables \ref{tab:annotation-prompt-1}--\ref{tab:annotation-prompt-3} contain the specific annotation prompt used.

\begin{table*}[t]
  \centering
  \small
  \setlength{\tabcolsep}{8pt}
  \renewcommand{\arraystretch}{1.15}
  \begin{tabular}{l p{0.5\textwidth} c c c}
    \toprule
    \textbf{Area} & \textbf{Conference} & \textbf{n} & \textbf{Year} & \textbf{Disclosure} \\
    \midrule
    \multirow{4}{*}{AI}
      & NeurIPS, ACL, EMNLP, SIGIR, WWW & 5 & 2026 & \cellcolor{highlightgreen}Yes \\
      & ICLR, NAACL                      & 2 & 2025 & \cellcolor{highlightgreen}Yes \\
      & AAAI, IJCAI, CVPR, ECCV, ICML    & 5 & 2026 & \cellcolor{highlightred}No   \\
      & ICCV                             & 1 & 2025 & \cellcolor{highlightred}No   \\
    \midrule
    \multirow{4}{*}{Systems}
      & ISCA, SIGCOMM, NSDI, CCS, IEEE S\&P, SIGMOD, DAC, RTAS, HPDC, ICS, MobiSys, IMC, SIGMETRICS, FSE, ICSE & 15 & 2026 & \cellcolor{highlightgreen}Yes \\
      & RTSS, SC                         & 2  & 2025 & \cellcolor{highlightgreen}Yes \\
      & ASPLOS, USENIX Security, VLDB, EMSOFT, MobiCom, SenSys, OSDI, SOSP, PLDI, POPL & 10 & 2026 & \cellcolor{highlightred}No \\
      & MICRO, ICCAD                     & 2  & 2025 & \cellcolor{highlightred}No   \\
    \midrule
    \multirow{3}{*}{Theory}
      & STOC, CRYPTO, EuroCrypt          & 3 & 2026 & \cellcolor{highlightgreen}Yes \\
      & SODA, CAV, LICS                  & 3 & 2026 & \cellcolor{highlightred}No   \\
      & FOCS                             & 1 & 2025 & \cellcolor{highlightred}No   \\
    \midrule
    \multirow{4}{*}{Interdisciplinary}
      & SIGGRAPH, SIGGRAPH Asia, SIGCSE, CHI, ICRA, VIS, VR & 7 & 2026 & \cellcolor{highlightgreen}Yes \\
      & IROS                             & 1 & 2025 & \cellcolor{highlightgreen}Yes \\
      & ISMB, RECOMB, Pervasive, UIST, RSS & 5 & 2026 & \cellcolor{highlightred}No \\
      & EC, WINE, UbiComp                & 3 & 2025 & \cellcolor{highlightred}No   \\
    \bottomrule
  \end{tabular}
  \caption{AI disclosure policy presence across major CS conferences.}
  \label{tab:conferences}
\end{table*}

\begin{table*}[t]
  \centering
  \small
  \setlength{\tabcolsep}{4pt}
  \begin{tabular}{l p{0.72\textwidth} r r}
    \toprule
    \textbf{Society} & \textbf{Conferences} & \textbf{n} & \textbf{\#Changes} \\
    \midrule
    AAAI & --                                                                 & 0  & 1 \\
    \addlinespace
    ACL  & ACL, EMNLP, NAACL                                                  & 3  & 2 \\
    \addlinespace
    ACM  & SIGIR, WWW, SIGCOMM, CCS, SIGMOD, DAC, HPDC, ICS, MobiSys, IMC, SIGMETRICS, FSE, ICSE, STOC, CRYPTO, EuroCrypt, SIGGRAPH, SIGGRAPH Asia, SIGCSE, CHI, ISCA & 21 & 1 \\
    \addlinespace
    IEEE & ISCA, IEEE S\&P, ICSE, ICRA, IROS, VIS, VR                         & 7  & 4 \\
    \addlinespace
    None & ICLR, NeurIPS, NSDI, RTAS, RTSS, SC                                & 6  & -- \\
    \bottomrule
  \end{tabular}
  \caption{Conferences inspired by each society's disclosure policy, and the number of revisions made.}
  \label{tab:policies}
\end{table*}

\begin{table*}[t]
  \centering
  \small
  \begin{tabular}{p{0.2\textwidth} p{0.7\textwidth}}
    \toprule
    \textbf{Phase} & \textbf{Task} \\
    \midrule
    \multirow{4}{=}{Idea Generation}
      & Discover research topics or identify gaps in current research \\
      & Identify relevant literature \\
      & Summarize or analyse existing literature \\
      & Propose new hypotheses \\
    \midrule
    \multirow{2}{=}{Research Design}
      & Help design research methodology or experiments \\
      & Help develop theoretical models or conceptual frameworks \\
    \midrule
    \multirow{5}{=}{Data Collection}
      & Suggest experimental parameters (e.g., choosing sample size, experimental conditions, or training settings) \\
      & Formulate questions for surveys or interviews \\
      & Transcribe recordings of research material (e.g. interviews, workshops or focus groups)\\
      & Generate synthetic data sets \\
      & Clean and reformat dataset \\
    \midrule
    \multirow{4}{=}{Data Analysis}
      & Create or edit software code for data analysis, statistical analysis or simulations \\
      & Support qualitative and thematic data analysis and coding \\
      & Help pattern recognition in data\\
      & Create or modify scientific figures or images \\
    \midrule
    \multirow{6}{=}{Writing and Reporting}
      & Suggest a structure for a research paper \\
      & Draft parts of a research paper \\
      & Propose a title or keywords for a research paper \\
      & Edit a research paper to improve readability and/or language \\
      & Format references \\
      & Translate a research paper into a different language \\
    \bottomrule
  \end{tabular}
  \caption{Taxonomy of research phases and associated tasks.}
  \label{tab:tasks}
\end{table*}

\begin{table*}[t]
  \centering
  \small
  \begin{tabular}{l p{0.25\textwidth} p{0.5\textwidth}}
    \toprule
    \textbf{Detail} & \textbf{Description} & \textbf{Example} \\
    \midrule
    Task
      & The specific task AI was used for.
      & ``In writing this work, LLMs have been used in \sethlcolor{highlight}\hl{helping find relevant works, format LaTeX, and implement code, and proofread.}'' \\
    \addlinespace
    \midrule
    Model name
      & The AI model or system that was used.
      & ``Specifically, we employed \sethlcolor{highlight}\hl{Claude Sonnet 4 (Anthropic) and GPT-5 (OpenAI)} for language polishing and refinement purposes.'' \\
    \addlinespace
    \midrule
    Reason
      & The reason AI was used instead of a non-AI method.
      & ``We use large language models (LLMs) \sethlcolor{highlight}\hl{to support labor-intensive and mistake-prone work.}'' \\
    \addlinespace
    \midrule
    Non-use
      & Tasks for which AI was not used
      & `` LLMs were \sethlcolor{highlight}\hl{not used for generating novel scientific ideas, experiments, or analyses.}'' \\
    \addlinespace
    \midrule
    Human oversight
      & The extent of human oversight or contributions during or after AI use.
      & ``All outputs from LLMs were \sethlcolor{highlight}\hl{carefully reviewed, verified, and edited by the authors} to ensure correctness and originality.'' \\
    \addlinespace
    \midrule
    Responsibility
      & The authors’ declaration of responsibility for outcomes of AI use, such as potential errors or biases.
      & ``The authors retain \sethlcolor{highlight}\hl{full responsibility for the entire content, including any errors or inaccuracies.}'' \\
    \addlinespace
    \midrule
    Purpose of disclosure
      & The purpose of the disclosure itself (why the information is shared), such as for transparency or compliance.
      & ``\sethlcolor{highlight}\hl{In accordance with ICLR guidelines}, we disclose that Large Language Models (LLMs) were used during the preparation of this manuscript.'' \\
    \addlinespace
    \bottomrule
  \end{tabular}
  \caption{Disclosure details, with the relevant portion of each example highlighted.}
  \label{tab:details}
\end{table*}

\begin{table*}[ht]
\centering
\begin{promptbox}{Welcome Screen}
\small
This survey aims to understand how readers and researchers perceive AI use in academic writing and research. You will be asked about your views on when and how AI use should be disclosed, and how such disclosures differ based on the research subtask and level of human involvement.\\

Your responses will help inform better guidelines for transparency and ethical AI use in research communication.\\

Participation is voluntary and anonymous, and the survey should take approximately \textbf{15–20} minutes to complete.
\end{promptbox}
\caption{Welcome screen of the survey instrument.}
\label{tab:survey-welcome}
\end{table*}

\begin{table*}[ht]
\centering
\begin{promptbox}{Introduction and Consent}
\small
\begin{enumerate}
    \item \textbf{Purpose and Goal} \\An AI Disclosure Statement in a research paper is a brief description of how Generative AI tools (such as ChatGPT, Gemini, and Claude) were used by the authors during the entire research process. The inclusion of such statements is being gradually mandated by academic publishers, individual journals, and conferences, however with no consensus on the presence or framing across institutions. The purpose of this survey is to understand the opinions of academic readers on the importance of the presence and framing of AI disclosure statements, as well as their necessity across various research tasks with differing levels of human involvement. This information will then be used to arrive at framing recommendations for AI Disclosure Statements, and an AI Disclosure Checklist that aims to inform policy regarding AI use and disclosure at academic and publishing organizations.
    \item \textbf{Procedure} \\After confirming the informed consent, you will be directed to answer a mixture of multiple-choice, Likert scale, and open-ended questions in the following sections: Presence and Framing of AI Disclosure Statements, Importance of AI Disclosure in Idea Generation, Research Design, Data Collection, Data Analysis, and Writing and Reporting. Then, you will be prompted to optionally share additional perspectives on the necessity of AI Disclosures and fill in some demographic information. Finally,  you will be redirected to a separate Google Form to provide your e-mail address to receive compensation. 
    \item \textbf{Participation and Compensation} \\Your participation in this user study is completely voluntary. Compensation is available for participation in this study. If you opt in to be compensated, you will receive a compensation of \$10/\euro 10/ \rupee 500, depending on your geographical region, in the form of an Amazon Gift Card. You may withdraw and discontinue participation at any time without penalty. Repeated participation in the survey is not permitted.
    \item \textbf{Confidentiality} \\All information gathered in this study is anonymous and will be kept completely confidential.  No reference will be made in written or oral materials that could link you to this study. All records will be stored securely at [RETRACTED INSTITUTE NAME]. 
    \item \textbf{Investigators} \\If you have any questions or concerns about this research, please feel free to contact [RETRACTED INVESTIGATOR DETAILS]
    \item \textbf{Informed Consent and Agreement} \\ \ding{111} I understand the explanation provided to me. I understand that this declaration of consent is revocable at any time. I have had all my questions answered to my satisfaction, and I voluntarily agree to participate in this survey.
\end{enumerate}
\end{promptbox}
\caption{Introduction and Consent section of the survey instrument.}
\label{tab:survey-intro}
\end{table*}

\begin{table*}[ht]
\centering
\begin{promptbox}{Presence and Framing of AI Disclosure Statements}
\small
\begin{enumerate}
    \item Indicate how strongly you agree or disagree with the following statements. \\\textit{Scale: $1$ = Strongly disagree, $2$ = Disagree, $3$ = Neither Agree nor Disagree, $4$ = Agree, $5$ = Strongly agree}
    \begin{enumerate}
        \item I usually notice AI-related disclosure statements in the research papers I read.
        \item I actively look for AI disclosure statements when reading research papers.
        \item It is important to me that I am aware of the extent of AI usage by the authors of a research paper.
        \item AI disclosure statements help me make an informed evaluation of the quality, credibility or originality of the research paper.
    \end{enumerate}
    \item \textit{[Depending on response to 1(c)]}
    \begin{enumerate}
        \item \textit{If >= $4$:} Why is it important for you to be aware of the extent of AI use by the authors of a research paper?
        \item \textit{If == $3$:} When is it important for you to be aware of the extent of AI use by the authors of a research paper?
        \item \textit{If <= $2$:} Why is it not important for you to be aware of the extent of AI use by the authors of a research paper?
    \end{enumerate}
    \item In your opinion, where should AI disclosure statements be located in a research paper, if at all? \\\ding{111} At the beginning of a research paper\\\ding{111} In the methodology section of a research paper\\\ding{111} In a separate section at the end of the research paper\\\ding{111} In the acknowledgments section of a research paper\\\ding{111} Throughout the paper, in each section where AI played a role\\\ding{111} Outside the main text (e.g., cover sheet, landing page, or supplementary materials)\\\ding{111} I do not believe AI disclosure statements are necessary
    \item Please describe the reasons behind your preferred locations for an AI disclosure statement. 
    \item In your opinion, what should an AI disclosure statement include, if present? \\\ding{111} The specific task AI was used for\\\ding{111} The AI model or system that was used\\\ding{111} The reason AI was used instead of a non-AI method\\\ding{111} Tasks for which AI was not used\\\ding{111} The extent of human oversight or contributions during or after AI use\\\ding{111} The authors’ declaration of responsibility for outcomes of AI use, such as potential errors or biases\\\ding{111} The purpose of the disclosure itself (why the information is shared), such as for transparency or compliance\\\ding{111} I do not believe AI disclosure statements are necessary
    \item In addition, what level of detail, if any, is appropriate for an AI disclosure statement in a research paper?\\\ding{109} One to two sentences\\\ding{109} A few sentences or one short paragraph\\\ding{109} Multiple paragraphs\\\ding{109} Length and detail proportionate to the extent of AI use (this could potentially span multiple paragraphs)\\\ding{109} I do not believe AI disclosure statements are necessary
\end{enumerate}
\end{promptbox}
\caption{Presence and Framing of AI Disclosure Statements section of the survey instrument.}
\label{tab:survey-presence-importance}
\end{table*}

\begin{table*}[ht]
\centering
\begin{promptbox}{Importance of AI Disclosure in Idea Generation}
\small
Before proceeding, please take a moment to review the tasks below. If you believe that none of the tasks below require AI disclosure, regardless of the level of human involvement, select the option below to skip this section. We'll record your answers as 'need not be disclosed.' If even one task or condition gives you pause, leave this unselected and answer the questions below instead.\\\ding{111} None of the tasks below require AI disclosure, regardless of human involvement.
\begin{enumerate}
    \item For each of the following idea generation tasks, please indicate how necessary you think AI disclosure is. \\\textit{Scale: $1$ = Need not be disclosed, $2$ = Disclosure is rarely needed, $3$ = Disclosure is sometimes needed, $4$ = Disclosure is usually needed, $5$ = Disclosure is always needed}
    \begin{enumerate}
        \item Discover research topics or identify gaps in current research
        \item Identify relevant literature
        \item Summarize or analyse existing literature
        \item Propose new hypotheses
    \end{enumerate}
    \item For each of the following idea generation tasks, if human involvement is high, please indicate how necessary you think AI disclosure is. \\\textit{High human involvement is when the author provides most of the contribution and initiative, while AI outputs are closely supervised and verified. \\Scale: $1$ = Need not be disclosed, $2$ = Disclosure is rarely needed, $3$ = Disclosure is sometimes needed, $4$ = Disclosure is usually needed, $5$ = Disclosure is always needed}
    \begin{enumerate}
        \item Discover research topics or identify gaps in current research
        \item Identify relevant literature
        \item Summarize or analyse existing literature
        \item Propose new hypotheses
    \end{enumerate}
    \item For each of the following idea generation tasks, if human involvement is low, please indicate how necessary you think AI disclosure is. \\\textit{Low human involvement is when the AI assistant provides most of the contribution and initiative, and AI outputs are loosely supervised and verified by the author. \\Scale: $1$ = Need not be disclosed, $2$ = Disclosure is rarely needed, $3$ = Disclosure is sometimes needed, $4$ = Disclosure is usually needed, $5$ = Disclosure is always needed}
    \begin{enumerate}
        \item Discover research topics or identify gaps in current research
        \item Identify relevant literature
        \item Summarize or analyse existing literature
        \item Propose new hypotheses
    \end{enumerate}
\end{enumerate}
\end{promptbox}
\caption{Importance of AI Disclosure in Idea Generation section of the survey instrument. All subsequent phases follow the same template, and are thus excluded for brevity.}
\label{tab:survey-phase}
\end{table*}

\begin{table*}[ht]
\centering
\begin{promptbox}{Additional Views on AI Disclosure Necessity}
\small
\begin{enumerate}
    \item Are there any other research tasks for which you think AI disclosure is necessary?
    \item In general, how do you think the level of human involvement affects whether AI use should be disclosed across different research tasks?
\end{enumerate}
\end{promptbox}
\caption{Additional Views on AI Disclosure Necessity section of the survey instrument.}
\label{tab:survey-additional-views}
\end{table*}

\begin{table*}[ht]
\centering
\begin{promptbox}{Demographics}
\small
\begin{enumerate}
    \item What is your age?
    \item What gender(s) do you identify with the most?\\\ding{111} Female \ding{111} Male \ding{111} Non-binary \ding{111} Prefer not to disclose
    \item Which country are you currently based in for your primary work or study?
    \item What is your current professional or academic affiliation?\\\ding{109} Academia \ding{109} Industry
    \item \textit{[Depending on response to 4]} What is your position at your current institution?
    \begin{enumerate}
        \item \textit{If Academia:} \\\ding{109} Undergraduate Student \ding{109} Master’s Student \ding{109} PhD Student \ding{109} Postdoctoral Researcher \\\ding{109} Research Associate \ding{109} Assistant Professor \ding{109} Associate Professor \ding{109} Professor\\\ding{109} Prefer not to answer \ding{109} Other
        \item \textit{If Industry:} \\\ding{109} Research Associate \ding{109} Junior Research Scientist \ding{109} Senior Research Scientist \\\ding{109} Principal Research Scientist \ding{109} Prefer not to answer \ding{109} Other
    \end{enumerate} 
    \item What is your major field of study/research?\\\ding{109} Artificial Intelligence\\\ding{109} Systems\\\ding{109} Theory\\\ding{109} Other / Interdisciplinary Areas
    \item How many years of research experience do you have?\\\ding{109} <$1$ year \ding{109} $1$-$3$ years \ding{109} $4$-$7$ years \ding{109} $8$-$10$ years \ding{109} >$10$ years
    \item How many years of experience do you have with reviewing research papers for peer review?\\\ding{109} <$1$ year \ding{109} $1$-$3$ years \ding{109} $4$-$7$ years \ding{109} $8$-$10$ years \ding{109} >$10$ years
    \item How frequently do you read research papers?\\\ding{109} Never \ding{109} Occasionally \ding{109} Monthly \ding{109} Weekly \ding{109} Daily
    \item In your research field, how common is the use of AI tools in research workflows?\\\textit{Scale: $1$ = Very uncommon, $2$ = Somewhat uncommon, $3$ = Neither common nor uncommon, $4$ = Somewhat common, $5$ = Very common}
    \item For each of the following research phases, please indicate whether you have used and/or disclosed AI assistance in any of your past research papers. \\\textit{We'd like to reiterate that responses are anonymous.\\Options: \ding{109} Didn't use AI assistance \ding{109} Used and disclosed AI assistance \ding{109} Used but didn't disclose AI assistance}
    \begin{enumerate}
        \item Idea Generation
        \item Research Design
        \item Data Collection
        \item Data Analysis
        \item Writing and Reporting
    \end{enumerate}
    \item Indicate how strongly you agree or disagree with the following statement: I am comfortable with the increasing integration of AI tools into research workflows.\\\textit{Scale: $1$ = Strongly disagree, $2$ = Disagreee, $3$ = Neither agree nor disagree, $4$ = Agree, $5$ = Strongly agree}
    \item \textit{[Depending on response to 12]}
    \begin{enumerate}
        \item \textit{If <=2:} What aspects of the increasing integration of AI tools into research workflows makes you uncomfortable?
        \item \textit{If ==3:} In what situations are you comfortable with the integration of AI tools into research workflows, and in what situations are you not?
        \item \textit{If >=4:} Are there some specific reasons behind your stance on the integration of AI tools into research workflows, or why you feel this way?
    \end{enumerate}
\end{enumerate}
\end{promptbox}
\caption{Demographics section of the survey instrument.}
\label{tab:survey-demographics-text}
\end{table*}

\begin{table*}[ht]
  \centering
  \small
  \setlength{\tabcolsep}{8pt}
  \renewcommand{\arraystretch}{1.15}
  \begin{tabular}{p{0.4\textwidth} p{0.45\textwidth} r}
    \toprule
    \textbf{Demographic} & \textbf{Level} & \textbf{$n$} \\
    \midrule
    \multirow{4}{*}{Gender}
      & Female                                     & $36$ \\
      & Male                                       & $67$ \\
      & Non-binary                 & $1$ \\
      & Prefer not to disclose                         & $5$ \\
    \midrule
    \multirow{6}{*}{Country of work / residence}
      & India                                     & $46$ \\
      & United States                             & $40$ \\
      & Germany                            & $7$ \\
      & Canada                                   & $5$ \\
      & United Arab Emirates                                     & $4$ \\
      & Israel                         & $1$ \\
    \midrule
    \multirow{4}{*}{Affiliation}
      & Academia                    & $104$ \\
      & Industry                                  & $3$ \\
      & Other                                     & $2$ \\
    \midrule
    \multirow{5}{*}{Field of study}
      & Artificial Intelligence        & $52$ \\
      & Systems                                   & $13$ \\
      & Theory                                    & $12$ \\
      & Other / Interdisciplinary Areas                         & $32$ \\
    \midrule
    \multirow{4}{*}{Years of research experience}
      & <$1$ years                                & $22$ \\
      & $1$--$3$ years                                & $48$ \\
      & $4$-$7$ years                               & $21$ \\
      & $8$-$10$ years                               & $11$ \\
      & >$10$ years                                 & $7$ \\
    \midrule
    \multirow{4}{*}{Years of review experience}
      & <$1$ years                                & $56$ \\
      & $1$--$3$ years                                & $29$ \\
      & $4$-$7$ years                               & $15$ \\
      & $8$-$10$ years                               & $5$ \\
      & >$10$ years                                 & $4$ \\
    \bottomrule
  \end{tabular}
  \caption{Demographics of survey respondents ($N=109$). Mean age: $26.64$ years (SD = $6.28$). Counts in the $n$ column are unweighted respondent totals; categories within each demographic are mutually exclusive.}
  \label{tab:survey-demographics}
\end{table*}

\begin{table*}[ht]
\centering
\begin{promptbox}{ICLR Disclosure Extraction Prompt}
\small
You are extracting AI/LLM usage disclosure statements from academic papers.\\

  Your task: find a dedicated section where the authors disclose how they used AI tools or
  Large Language Models (LLMs) in **preparing or writing the manuscript** — for example,
  grammar checking, text polishing, writing assistance, or code editing.\\

  IMPORTANT DISTINCTIONS:
  - DO extract: sections about LLMs used for manuscript preparation (writing, editing, grammar,
    coding assistance for the paper itself).
  - DO extract: negative disclosures where authors explicitly state they did NOT use LLMs.
  - DO NOT extract: inline mentions of LLM use embedded within Methods, Experiments,
    or Results sections that are not part of a dedicated disclosure statement or paragraph.\\

  Where to look:
  - This section almost always appears near the END of the paper (after Conclusions, near
    Acknowledgments or References).
  - Common headings include: "LLM Usage", "Use of Large Language Models", "AI Disclosure",
    "Declaration of LLM Usage", "LLM Usage Statement", etc.
  - It may also be embedded inside an "Ethics Statement" or "Acknowledgments" section.
  - Occasionally there is no heading — look for a standalone paragraph near the end of the
    paper that explicitly addresses AI tool usage in manuscript preparation.\\

  Return a JSON object with:
  - "found": true if a disclosure section or paragraph was found, false otherwise
  - "heading": the exact heading of the section (empty string if embedded or no heading)
  - "content": the full text content of the disclosure (empty string if not found)\\

  Return ONLY valid JSON.\end{promptbox}
\caption{Prompt used for extracting ICLR disclosures.}
\label{tab:extraction-prompt}
\end{table*}

\begin{table*}[ht]
\centering
\begin{promptbox}{EMNLP Disclosure Extraction Prompt 1}
\small
You are reading the "Responsible NLP Checklist" of an EMNLP paper.\\

The checklist has a legend at the top that explains the symbols:
  \ding{51} (checkmark / tick)  = the authors responded YES
  \ding{55} (cross / X)         = the authors responded NO
  N/A                   = does not apply
  empty box             = no response\\

These two symbols are DIFFERENT and must not be confused:
  \ding{51} = YES (a tick, checkmark, or filled checkmark — the answer is affirmative)
  \ding{55} = NO  (a cross, X, or filled X — the answer is negative)\\

Find section E, which concerns AI assistant use:
  E.  Did you use AI assistants (e.g., ChatGPT, Copilot) in your research, coding, or writing?
  E1. If you used AI assistants, did you include information about their use?\\

Your task:\\

1. Look at the symbol next to question E (not E1). Read it carefully.
   - If it is \ding{55} (cross/X/NO) or N/A: set ai\_used = false and stop.
   - If it is \ding{51} (tick/checkmark/YES): set ai\_used = true and continue.\\

2. Read the answer text written next to or below question E1 (do NOT rely on E1's symbol —
   it varies across submissions and is unreliable). Instead, classify by the text itself:\\

   - "paper\_section": the text is a location pointer — either a bare pointer (just a section
     number, heading name, appendix letter, or page, e.g. "Section 5", "Ethics Statement",
     "Appendix A", "page 8", "see §4") OR a wrapper sentence that only says the disclosure IS
     in a section without describing WHAT AI was used for (e.g. "We include information in the
     Acknowledgement section", "We disclosed AI usage in Appendix G", "In the Ethics Statement
     we specified the AI usage", "Yes we include the use of AI assistants in Section 3",
     "We mention the usage in pre-appendix", "We have detailed how to use it in Appx. G").
     Key test: does the text tell you WHAT AI was used for? If no — it is paper\_section.
     → set disclosure\_location = "paper\_section", copy the text to section\_reference.\\

   - "inline": the text describes WHAT AI was used for — it contains actual content about the
     usage, not just a pointer to where it is described. It may also mention a section, but the
     description itself is informative without needing to look elsewhere
     (e.g. "Used for grammar correction", "We used ChatGPT only for writing polish",
     "Used for machine translation. We outline in Section 4.6.").
     → set disclosure\_location = "inline", copy the full text to content.\\

   - "none": E1 has no meaningful answer text (blank, just "N/A", bare "yes"/"no").
     → set disclosure\_location = "none".\\

Return a JSON object with exactly these fields:
- "ai\_used": true or false
- "disclosure\_location": one of "inline", "paper\_section", "none"
- "content": the inline disclosure text if disclosure\_location is "inline", else ""
- "section\_reference": the section/page pointer if disclosure\_location is "paper\_section", else ""
- "e\_symbol\_seen": the exact symbol you saw next to question E (e.g. "\ding{51}", "\ding{55}", "N/A", "empty")\\

Return ONLY valid JSON.
\end{promptbox}
\caption{First-stage prompt for the two-stage EMNLP extraction pipeline. Reads the Responsible NLP Checklist (Section E and E1) and returns AI-use status, disclosure type (inline vs.\ paper-section pointer), and the corresponding content or section reference.}
\label{tab:emnlp-loc-extraction-prompt}
\end{table*}

\begin{table*}[ht]
\centering
\begin{promptbox}{EMNLP Disclosure Extraction Prompt 2}
\small
You are extracting an AI usage disclosure from an EMNLP paper.\\

The authors' checklist section E1 points to this location:
"{section\_reference}"\\

Locate that section, appendix, or page. Your task is to determine whether it contains a
GENUINE AI usage disclosure and, if so, extract ONLY the AI disclosure portion.\\

A GENUINE disclosure is one where the authors explicitly address their use (or non-use) of
AI tools in the context of THIS paper — for example:
  - Using LLMs or AI assistants to write, edit, polish, proofread, or translate the manuscript
  - Using AI coding assistants (Copilot, Cursor) to write code for the paper
  - Explicitly stating they did NOT use any AI tools
  - Any deliberate statement of AI usage made to inform readers about how the paper was produced\\

NOT genuine (set found = false):
  - Sections that only describe LLMs as the subject of study or research methodology
    (e.g. "we used GPT-4 to generate our dataset", "we evaluated ChatGPT on our benchmark")
    with no separate statement about manuscript preparation
  - Generic ethics or limitations sections with no mention of AI tool use at all
  - Sections that do not exist in the paper\\

IMPORTANT — partial extraction:
  The referenced section may be a broader section (Ethics Statement, Acknowledgments,
  Limitations, etc.) that contains other content unrelated to AI tool usage. In that case,
  extract ONLY the sentences or paragraph that specifically address AI/LLM usage.
  Do NOT copy the entire section — only the AI disclosure part.
  Set "heading" to the section's actual heading even if only part of it is extracted.\\

Return a JSON object with:
- "found": true if a genuine AI usage disclosure was found, false otherwise
- "heading": the exact section heading where the disclosure appears (empty string if no heading)
- "content": ONLY the AI disclosure text, not the full surrounding section\\

Return ONLY valid JSON.
\end{promptbox}
\caption{Second-stage prompt for the two-stage EMNLP extraction pipeline. Reads the referenced section of the paper PDF and returns whether a genuine AI usage disclosure is present, the section heading, and the disclosure text itself.}
\label{tab:emnlp-extraction-prompt}
\end{table*}

\begin{table*}[ht]
  \centering
  \small
  \setlength{\tabcolsep}{8pt}
  \renewcommand{\arraystretch}{1.2}
  \begin{tabular}{l rrr rrr r}
    \toprule
    \multirow{2}{*}{\textbf{Annotator}}
      & \multicolumn{3}{c}{\textbf{Details}}
      & \multicolumn{3}{c}{\textbf{Tasks}}
      & \multirow{2}{*}{\textbf{Length Acc.}} \\
    \cmidrule(lr){2-4} \cmidrule(lr){5-7}
      & \textbf{P} & \textbf{R} & \textbf{F1}
      & \textbf{P} & \textbf{R} & \textbf{F1}
      & \\
    \midrule
    Annotator $1$   & $98.2$ & $97.4$ & $97.8$ & $91.8$ & $93.7$ & $92.7$ & $90.0$ \\
    Annotator $2$    & $98.2$ & $95.7$ & $96.9$ & $88.4$ & $90.2$ & $89.3$ & $86.0$ \\
    Annotator $3$  & $91.1$ & $98.6$ & $94.7$ & $91.1$ & $88.7$ & $89.9$ & $87.0$ \\
    \midrule
    \textit{Average} & \textit{95.8} & \textit{97.2} & \textit{96.5} & \textit{90.4} & \textit{90.9} & \textit{90.6} & \textit{87.7} \\
    \bottomrule
  \end{tabular}
  \caption{Validation of LLM-based annotation against $3$ human annotators on $N=100$ disclosures. We report micro precision (P), recall (R), and F1 for the multi-label \textbf{Details} and \textbf{Tasks} schemas, and classification accuracy for the single-label \textbf{Length} schema. All values are percentages; numbers compare each human annotator's labels to Gemini's annotations.}
  \label{tab:annotation-validation}
\end{table*}

\begin{table*}[ht]
  \centering
  \small
  \setlength{\tabcolsep}{6pt}
  \renewcommand{\arraystretch}{1.2}
  \begin{tabular}{l l rrr rrr r}
    \toprule
    \multirow{2}{*}{\textbf{Annotator A}}
      & \multirow{2}{*}{\textbf{Annotator B}}
      & \multicolumn{3}{c}{\textbf{Details}}
      & \multicolumn{3}{c}{\textbf{Tasks}}
      & \multirow{2}{*}{\textbf{Length Acc.}} \\
    \cmidrule(lr){3-5} \cmidrule(lr){6-8}
      & & \textbf{P} & \textbf{R} & \textbf{F1}
        & \textbf{P} & \textbf{R} & \textbf{F1}
        & \\
    \midrule
    Annotator $1$ & Annotator $2$    & $94.8$ & $96.5$ & $95.6$ & $90.2$ & $90.2$ & $90.2$ & $81.0$ \\
    Annotator $1$ & Annotator $3$  & $98.1$ & $89.9$ & $93.8$ & $87.3$ & $91.6$ & $89.4$ & $90.0$ \\
    Annotator $2$  & Annotator $3$  & $98.6$ & $88.7$ & $93.4$ & $84.7$ & $88.8$ & $86.7$ & $80.0$ \\
    \midrule
    \textit{Average} & & \textit{97.2} & \textit{91.7} & \textit{94.3} & \textit{87.4} & \textit{90.2} & \textit{88.8} & \textit{83.7} \\
    \bottomrule
  \end{tabular}
  \caption{Pairwise inter-annotator agreement on $N=100$ disclosures. We report micro precision (P), recall (R), and F1 for the multi-label \textbf{Details} and \textbf{Tasks} schemas, and classification accuracy for the single-label \textbf{Length} schema. All values are percentages.}
  \label{tab:iaa}
\end{table*}

\begin{table*}[ht]
\centering
\begin{promptbox}{Disclosure Annotation Prompt ($1$/$3$)}
\small
You are an expert annotator of AI usage disclosure statements in academic papers.\\

Given a disclosure statement, return a JSON object with exactly these three keys:\\

"details" — which types of information the disclosure contains.
Include only keys that are explicitly present; do not infer.\\

  "task": The disclosure names at least one specific task AI was used for.\\

  "model": A specific AI model or tool is named (e.g. ChatGPT, GPT-4, Claude, Gemini, \\Copilot). NOT for generic references like 'an LLM'.\\

  "reason": The disclosure explicitly states WHY AI was used instead of a non-AI approach — a justification clause, not just a description of what was done. Rare.\\

  "non\_use": The disclosure explicitly states tasks for which AI was NOT used.\\

  "human\_oversight": Authors explicitly state their own independent contributions ("all research ideas were ours", "all analysis was done by the authors"), OR explicitly describe reviewing/verifying/editing AI outputs before inclusion.\\

  "responsibility": Authors make a formal declaration of accountability for the content (e.g. 'take full responsibility', 'authors are responsible for').\\

  "purpose": The disclosure explicitly cites a reason for making the disclosure itself — conference policy, ethics guidelines, compliance.\\

Disambiguation:\\
- "model" requires a specific name (ChatGPT, GPT-4, Gemini 2.5, Copilot, Grammarly…). Generic phrases like "a large language model" or "an LLM" do NOT trigger this. If a specific model is named only in the context of what was NOT used (e.g. "we did not use GPT or DeepSeek"), do NOT label "model".\\

- "human\_oversight" vs "responsibility": oversight requires an EXPLICIT statement — either about what the authors themselves contributed ("all research ideas were ours", "all analysis was done by the authors") or about reviewing/verifying AI outputs ("we reviewed all LLM-generated text", "authors verified and edited all suggestions"). Do NOT infer oversight from the fact that AI was used only for minor tasks (e.g. "we only used LLMs for writing polish" alone does not imply human\_oversight). "Taking full responsibility" is responsibility, not oversight. Both can appear together.\\

- "non\_use" requires an explicit negation statement ("not used for", "were not involved in", "did not use AI for"). Saying AI was "only" or "solely" used for X does NOT by itself imply non\_use — it just scopes the use. Only label non\_use when there is an explicit statement that AI was NOT used for something (e.g. "LLMs were not used for data analysis", "AI was not involved in experiment design"). Phrases like "all research ideas/contributions are our own" or "all analysis was done by the authors" are authorship claims (human\_oversight), NOT non\_use.\\

- "purpose" requires citing a specific external trigger for the disclosure itself ("in accordance with ICLR 2026 policy", "as required by the Code of Ethics"). Do not infer purpose from the existence of the disclosure alone.\\

- "reason" is rare — only include if the disclosure gives a forward-looking justification for WHY AI was chosen (e.g. "we used LLMs because manual annotation at this scale was not feasible", "AI was used to accelerate repetitive processing"). Statements about what AI did or did not affect ("these uses do not influence our results", "LLMs were not involved in the core methodology") are scope/impact statements, not reasons.\\
\end{promptbox}
\caption{Annotation prompt ($1$/$3$): defines the seven detail labels (\textit{task}, \textit{model}, \textit{reason}, \textit{non\_use}, \textit{human\_oversight}, \textit{responsibility}, \textit{purpose}) and the disambiguation rules.}
\label{tab:annotation-prompt-1}
\end{table*}

\begin{table*}[ht]
\centering
\begin{promptbox}{Disclosure Annotation Prompt ($2$/$3$)}
\small
"tasks" — which specific tasks AI was used FOR.
Include only tasks the disclosure explicitly mentions AI performing.\\

  "edit\_clarity": Edit, polish, or improve EXISTING text for grammar, clarity, style, or readability. Includes proofreading, sentence refinement, spell-checking, paraphrasing. Most common task.\\
  "draft\_paper": Generate NEW text or draft sections/paragraphs from scratch. Distinct from edit\_clarity: drafting = creating new content; editing = improving existing content.\\
  "paper\_structure": Suggest or plan the structure or outline of the paper, OR adjust the layout and presentation of existing figures, tables, or sections (e.g. "suggesting formatting adjustments for tables", "adjusting the layout of figures and tables", "restructuring sections for logical flow").\\
  "title\_keywords": Propose a title or keywords for the paper.\\
  "translate": Translate text between languages.\\
  "format\_references": Format or manage citations and references.\\
  "discover\_topics": Identify research gaps or brainstorm research directions.\\
  "identify\_literature": Search for, find, or suggest relevant papers and references.\\
  "summarize\_literature": Summarize or synthesize existing papers.\\
  "propose\_hypotheses": Propose new research hypotheses.\\
  "design\_methodology": Help design research methodology or experiments.\\
  "develop\_models": Help develop theoretical models or conceptual frameworks.\\
  "suggest\_parameters": Suggest experimental parameters or hyperparameters.\\
  "formulate\_questions": Formulate questions for surveys or interviews.\\
  "synthetic\_data": Generate synthetic or artificial datasets.\\
  "clean\_data": Clean, reformat, or preprocess datasets.\\
  "transcribe": Transcribe recordings or research material.\\
  "qualitative\_analysis": Support qualitative coding, thematic analysis, or annotation.\\
  "pattern\_recognition": Help detect patterns in data.\\
  "software\_code": Write, edit, debug, or complete code for analysis, simulations, or document preparation. Includes writing or editing LaTeX commands/macros/syntax, plotting scripts, and code assistants like Copilot or Cursor. Does NOT include adjusting the visual layout of figures and tables (use "paper\_structure") or general text formatting (use "edit\_clarity").\\
  "figures\_images": Generate or create new scientific figures, plots, diagrams, schematics, or visual content. Use this when AI produces the visual content itself — not when it adjusts the layout or presentation of existing figures/tables (use "paper\_structure" for that).\\
  "other": AI was clearly used for a task not covered above. Use sparingly — only when the use case is unambiguous but genuinely outside all categories (e.g. used as automated judge in experiments, acting as a research subject).\\

Disambiguation:\\
- "edit\_clarity" vs "draft\_paper": if the text says AI "polished", "refined", "improved", "corrected", "proofread" → edit\_clarity. If it says AI "wrote", "drafted", "rewrote sections", "generated initial text", "produced paragraphs" → draft\_paper. A disclosure can include both.\\

- "figures\_images" vs "paper\_structure": use "figures\_images" when AI generates or creates new visual content (plots, diagrams, schematics, figure captions as text). Use "paper\_structure" when AI adjusts, reformats, or suggests changes to the layout or presentation of existing figures and tables (e.g. "formatting adjustments for tables and figures", "adjusting the layout and presentation of figures and tables").\\

- "software\_code" vs "edit\_clarity": LaTeX command writing/editing (macros, environments, syntax) → software\_code. General wording like "LaTeX formatting" or "LaTeX styling" that refers to how text looks (not actual code) → edit\_clarity. When in doubt, if the phrase implies writing or correcting actual code/commands → software\_code; if it just means making the text look right → edit\_clarity.\\

- "identify\_literature" = searching for / finding papers. "summarize\_literature" = synthesizing or summarizing papers already found.\\

- "other": use only when AI is clearly used for something that fits no category above, e.g. acting as an automated evaluator/judge in experiments.\\
\end{promptbox}
\caption{Annotation prompt ($2$/$3$): defines the task labels (e.g., \textit{edit\_clarity}, \textit{draft\_paper}, \textit{software\_code}, \textit{synthetic\_data}) and disambiguation rules for borderline cases.}
\label{tab:annotation-prompt-2}

\end{table*}

\begin{table*}[ht]
\centering
\begin{promptbox}{Disclosure Annotation Prompt ($3$/$3$)}
\small
"length" — total length of the AI disclosure portion, exactly one of:\\
  "1-2 sentences"             — exactly 1 or 2 sentences. Count carefully: if there are 3 or more sentences, this label does not apply.\\
  "few sentences/1 paragraph" — exactly 3, 4, or 5 sentences, or one coherent paragraph of similar length\\
  "many paragraphs"           — 2 or more paragraphs, OR a bulleted/numbered list with 3 or more distinct items, OR 6 or more sentences\\
Count sentences explicitly before deciding.\\

Rules:\\
- Only annotate what is explicitly stated — do not infer or extrapolate.\\
- For tasks: omit any task not directly described as something AI was used for.\\
- If the disclosure only states what AI was NOT used for and describes no actual use, "tasks" must be empty and "task" must not appear in "details".\\
- Return empty lists [] if nothing applies.\\
- Return ONLY valid JSON, no explanation.\\
- The disclosure may be embedded within a broader section (e.g. ethics statement, acknowledgments, limitations) that contains unrelated content. Focus only on the portion that describes AI/LLM usage. In particular, "length" should reflect only the length of the AI disclosure portion, not the entire surrounding section.\\

EXAMPLES\\

Example 1 — short, editing only, explicit authorship claim:\\
Disclosure: "We used LLMs to assist with grammar and writing polishing. All equations, analysis, and research contributions are our own."
Output: {"details": ["task", "human\_oversight"], "tasks": ["edit\_clarity"], "length": "1-2 sentences"}\\

Example 2 — one paragraph, multiple details including explicit oversight:\\
Disclosure: "We used LLMs (specifically, OpenAI GPT-4.1, GPT-5 and Google Gemini 2.5) solely for checking grammar errors and improving the readability of the manuscript. The LLMs were not involved in research ideation, the development of research contributions, experiment design, data analysis, or interpretation of results. All substantive content and scientific claims were created entirely by the authors. The authors have reviewed all LLM-assisted text to ensure accuracy and originality, and take full responsibility for the contents of the paper."
Output: {"details": ["task", "model", "non\_use", "human\_oversight", "responsibility"], "tasks": ["edit\_clarity"], "length": "few sentences/1 paragraph"}\\

Example 3 — multiple tasks, explicit authorship claim:\\
Disclosure: "Large language models (LLMs) were used to revise sentences and correct grammar, to generate visualization code for some figures, and to assist with the implementation of the MULTIMNIST dataset. All conceptual contributions, experiment design, analysis, and the writing were done by the authors."
Output: {"details": ["task", "human\_oversight"], "tasks": ["edit\_clarity", "software\_code", "figures\_images"], "length": "few sentences/1 paragraph"}\\

Example 4 — drafting + editing + compliance purpose + responsibility, no explicit review of outputs:\\
Disclosure: "In this work, large language models served as an assistive tool in the research and writing process. The model's contributions included assisting with drafting and iteratively revising the manuscript — such as generating initial text and rephrasing sentences to improve clarity — and helping to write and debug Python code for the data analysis pipeline. In accordance with ICLR 2026 policy, we disclose this use. The authors take full responsibility for all content."
Output: {"details": ["task", "responsibility", "purpose"], "tasks": ["draft\_paper", "edit\_clarity", "software\_code"], "length": "few sentences/1 paragraph"}\\

Example 5 — pure non-use, no actual use described (tasks must be empty):\\
Disclosure: "This paper does not use LLMs for research ideation or paper writing."
Output: {"details": ["non\_use"], "tasks": [], "length": "1-2 sentences"}\\

Example 6 — model named only in non-use context (do NOT label "model"):\\
Disclosure: "This work does not involve the use of any large language models (LLMs), such as GPT, DeepSeek, or similar models."
Output: {"details": ["non\_use"], "tasks": [], "length": "1-2 sentences"}\\

Example 7 — explicit human oversight (review process described):\\
Disclosure: "LLMs were extensively used as annotation helpers, as analytical assistants for reasoning trace analysis, and as automated judges. They contributed to the writing process by generating initial drafts and assisting with revision. All LLM-generated content underwent rigorous human review and validation, with human authors verifying analyses and thoroughly editing all contributions. While LLMs served as powerful assistive tools, all final decisions were made by human authors."
Output: {"details": ["task", "human\_oversight"], "tasks": ["qualitative\_analysis", "draft\_paper", "edit\_clarity", "other"], "length": "many paragraphs"}
\end{promptbox}
\caption{Annotation prompt ($3$/$3$): defines the length categories, general output rules, and worked examples for the model to follow.}
\label{tab:annotation-prompt-3}
\end{table*}

\end{document}